\documentclass[sn-mathphys,Numbered]{sn-jnl}% Math and Physical Sciences Reference Style
\usepackage{graphicx}%
\usepackage{multirow}%
\usepackage{amsmath,amssymb,amsfonts}%
\usepackage{amsthm}%
\usepackage{mathrsfs}%
\usepackage[title]{appendix}%
\usepackage{xcolor}%
\usepackage{textcomp}%
\usepackage{manyfoot}%
\usepackage{booktabs}%
\usepackage{algorithm}%
\usepackage{algorithmicx}%
\usepackage{algpseudocode}%
\usepackage{listings}%
\usepackage{braket}
\usepackage{epstopdf}
\usepackage{amsfonts}
\usepackage{amssymb}
\usepackage{amsmath}
\usepackage{subfigure}
\theoremstyle{thmstyleone}%
\theoremstyle{thmstyletwo}%

\theoremstyle{thmstylethree}%

\begin{document}

\title[Article Title]{Hamiltonian formulation of linear non-Hermitian systems}

 %FamilyName	-> \sur{Ploeg}

\author*[1,2]{\fnm{Qi} \sur{Zhang}}\email{zhangqi0446@sina.com}

\affil*[1]{\orgdiv{College of Science}, \orgname{Liaoning Petrochemical University}, \orgaddress{\street{No.1 Dandong Road}, \city{Fushun}, \postcode{113001}, \state{Liaoning}, \country{People's Republic of China}}}

\affil[2]{\orgdiv{Liaoning Provincial Key Laboratory of Novel Micro-Nano Functional Materials}, \orgname{Liaoning Petrochemical University}, \orgaddress{\street{No.1 Dandong Road}, \city{Fushun}, \postcode{113001}, \state{Liaoning}, \country{People's Republic of China}}}


\abstract{For a linear non-Hermitian system, I demonstrate that a Hamiltonian can be constructed such that the non-Hermitian equations can be expressed exactly in the form of Hamilton's canonical equations. This is first shown for discrete systems and then extended to continuous systems. With this Hamiltonian formulation, I am able to identify a conserved charge by applying Noether's theorem and recognize adiabatic invariants. When applied to Hermitian systems, all the results reduce to the familiar ones associated with the Schr\"odinger equation.}




\maketitle

\section{Introduction}
For a conserved system, its Newton's equations of motion can be recast in the canonical Hamiltonian form as discussed extensively in textbooks on classical mechanics~\cite{Arnold}. Interestingly,
a Schr\"odinger equation can also be formulated in a Hamiltonian framework~\cite{Heslot,Weinberg}.
Consider a Schr\"odinger equation,
\begin{equation} \label{Schro0}
\text{i}\hbar\frac{\partial}{\partial t}|\psi\rangle=H_0|\psi\rangle\,,
\end{equation}
where $H_0$ is a Hermitian matrix and $|\psi\rangle = (\psi_1, \psi_2, ..., \psi_n)^T$ is a column vector.
It has been shown  that it  can be reformulated as Hamilton's equations of motion with ${\mathcal H}_0=\braket{\Psi|H_0|\Psi}$ as its  Hamiltonian and Poisson brackets between $\psi^*_i$'s and $\psi^*_j$'s~\cite{Heslot,Weinberg}.  Such a reformulation not only provides a distinct theoretical perspective into quantum dynamics described by the Schr\"odinger equation,
with access to various tools developed for Hamiltonian systems~\cite{Arnold,adia,Noether},
but also enables approximate quantum dynamics semi-classically~\cite{semi}.

In recent decades, sparked by studies of PT-symmetric systems~\cite{Bender} and bosonic Bogoliubov excitations~\cite{njp}, there has been substantial interest and research into non-Hermitian systems, both theoretically~\cite{Bender2,Longhi,West,Wang,chen,Fu,An,
Pi,Song,Xu,Li,Yao,zhangyang,Kou,YiZhang} and experimentally~\cite{e6,e9,e12,J1,J3,Guo,Xiao}.
In this work, I demonstrate that a Hamiltonian formulation can also be constructed for any linear non-Hermitian system. Specifically, when the Hermitian matrix $H_0$ in the Schr\"odinger equation
is replaced by a non-Hermitian matrix $h$ that
is diagonalizable, I show how to build a Hamiltonian such that the non-Hermitian equation can be derived as its equations of motion.  I first consider the case where $h$ is a finite matrix, and then extend the discussion to continuous systems. While specific non-Hermitian systems have been formulated previously using Hamiltonian or Lagrangian approaches~\cite{Rego,Liu2}, the formulation presented here is generic and applicable to any linear non-Hermitian system whose Hamiltonian matrix is diagonalizable.

This Hamiltonian formulation is applied to identify adiabatic invariants for non-Hermitian systems and find a conserved charge using Noether's theorem~\cite{Noether}. It is anticipated that more interesting results will emerge for linear non-Hermitian systems, as many powerful analytical tools have been developed for Hamiltonian systems~\cite{Arnold}. Just as many nonlinear Hermitian systems, such as
Newton's equations of motion and nonlinear Gross-Pitaevskii equations, can be cast into
canonical Hamiltonian form~\cite{Liu}, it is desirable to find a Hamiltonian formulation for nonlinear non-Hermitian systems. The method introduced here is limited to certain special cases, and a more comprehensive treatment of nonlinear non-Hermitian Hamiltonian systems remains an open challenge for future work.

\section{Discrete linear non-Hermitian systems}
\subsection{Key aspects of the dynamics}
Conventional quantum systems can be either discrete, such as spins, or continuous, such as electrons in atoms.
For clarity, we first consider a discrete linear non-Hermitian system, whose dynamical equation takes the form
\begin{equation} \label{Schro}
\text{i}\hbar\frac{\partial}{\partial t}|\psi\rangle=h|\psi\rangle\,.
\end{equation}
where, to distinguish from the Hermitian case, $h$ is denoted as an $n\times n$ non-Hermitian matrix that is diagonalizable and
$|\psi\rangle = (\psi_1, \psi_2, ..., \psi_n)^T$ is a vector in an $n$-dimensional linear space.
This serves as a discrete model before extending the formulation to continuous non-Hermitian systems.
In general,
the diagonalizable matrix $h$  has two sets of $n$ eigenvectors $|a_j\rangle$ and $|b_j\rangle$ with $j=1,\ldots n$ satisfying~\cite{F3},
\begin{equation} \label{right-left}
h|a_j\rangle=E_j|a_j\rangle,  \quad
\langle b_j|h=\langle b_j|E_j.
\end{equation}
They are biorthonormal,
\begin{equation} \label{right-left1}
\langle b_i|a_j\rangle=\delta_{ij},
\end{equation}
and complete,
\begin{equation}  \label{right-left2}
\sum_j |a_j\rangle\langle b_j|=1\,.
\end{equation}
The eigenvalues $E_j$ may be complex in general. The vectors $|a_j\rangle$ and $|b_j\rangle$ are referred to as the right and left eigenvectors of $h$, respectively. Since $h$ is non-Hermitian, satisfying $h^\dag\neq h$, the right and left eigenvectors are usually distinct, i.e. $|a_j\rangle\neq|b_j\rangle$.

This biorthogonal eigenbasis provides a framework to analyze the non-Hermitian dynamics. For example, consider a general state $|\psi(t)\rangle$ satisfying dynamics~(\ref{Schro}), which can be expanded as (when $h$ is diagonalizable such that Eq.~(\ref{right-left2}) holds),
\begin{equation}  \label{6}
|\psi(t)\rangle=\sum_{j}c_j(t)|a_j\rangle,
\end{equation}
where the coefficients
\begin{equation} \label{EV1}
c_j(t)=c_j(0) e^{-\frac{\text{i}}{\hbar}E_jt} =\langle b_j|\psi(t)\rangle,
\end{equation}
are given by the biorthonormal equation (\ref{right-left1}). Focusing on the subset of eigenstates with nonzero $c_j$, we can construct corresponding coefficients,
\begin{equation} \label{relationI}
\bar{c}_j(t)c_j(t)=|C_j|^2,
\end{equation}
where the $|C_j|^2$ are arbitrary constants. The variables $\bar{c}_j$ apparently evolve as
\begin{equation} \label{EV2}
\bar{c}_j(t)=\bar{c}_j(0)e^{\frac{\text{i}}{\hbar}E_jt},
\end{equation}
This defines a conjugate left state
\begin{equation} \label{MOMEN}
\langle\bar{\phi}(t)|=\sum_j \bar{c}_j(t) \langle b_j|,
\end{equation}
which, by the eigenrelation in Eq.~(\ref{right-left}), satisfies the conjugate dynamics
\begin{equation} \label{Schro2}
-\text{i}\hbar\frac{\partial}{\partial t}\langle\bar{\phi}|=\langle\bar{\phi}|h,
\end{equation}
showing $|\bar{\phi}\rangle$ evolves under the adjoint Hamiltonian $h^\dag$.

The dynamics given by Eq.~(\ref{Schro}) and (\ref{Schro2}) lead directly to
\begin{equation} \label{conserv}
\frac{\partial}{\partial t}\langle\bar{\phi}|\psi\rangle=0
\end{equation}
This shows that the overlap between the left and right states, $\langle\bar{\phi}|\psi\rangle$, is conserved over time. However, the norm of the right eigenstate $\langle\psi|\psi\rangle$ is in general not a constant due to the non-Hermiticity of the Hamiltonian operator $h$~\cite{Ryu}.

When the eigenvalues $E_j$ of the Hamiltonian $h$ are real, it is convenient to initialize the biorthogonal wavefunctions as
\begin{equation}
|\bar{c}_j(0)|=|c_j(0)|,
\end{equation}
for which $|\bar{c}_j(t)|=|c_j(t)|=|C_j|$ at any subsequent time t. In this case, the biorthogonal formulation retains the normalization condition at all times.
Additionally, when $h$ becomes Hermitian, the left and right eigenvectors become identical, i.e. $|a_j\rangle=|b_j\rangle$. Consequently, the biorthogonal wavefunction $\langle\bar{\phi}|$ reduces to the standard single wavefunction $\langle\psi|$, restoring the conventional form of the Schr\"odinger equation. Here, conservation of the overlap $\langle\bar{\phi}|\psi\rangle$ corresponds to conservation of the total probability $\langle\psi|\psi\rangle$, as in standard quantum mechanics.
The biorthogonal framework therefore provides a natural generalization of the Hermitian theory that recovers the standard results when $h$ is Hermitian.

\subsection{The Hamiltonian formulation}

The dynamics described by Eqs. (\ref{Schro}) and (\ref{Schro2}) allow establishing a Hamiltonian formulation for non-Hermitian systems in a straightforward manner. In this formulation, the state should be represented by canonical variables. We can naturally take $q_j = i\hbar\psi_j$ as the canonical variables.
Unlike the Hermitian case, since $h \neq h^\dagger$ here, the pairs $(i\hbar\psi_j, \psi_j^*)$ can no longer constitute canonical variables~\cite{Rego,Liu2}. This arises from the non-Hermiticity of $h$, which prevents $\psi_j$ and $\psi_j^*$ from satisfying the required Poisson bracket relations to be canonically conjugate. The non-Hermiticity induces additional constraints between the wavefunction and its complex conjugate, requiring a more careful construction of the phase space. Identifying an appropriate set of canonical variables is key to formulating a valid Hamiltonian framework for non-Hermitian systems.

Fortuitously, the right and left eigenstate dynamics in Eqs.~(\ref{Schro}) and (\ref{Schro2}) indicate that the components of the conjugate left state $\langle\bar{\phi}|=(\bar{\phi}_1,\bar{\phi}_2,\ldots,\bar{\phi}_n)$
constitute the canonical momenta conjugate to the canonical coordinates given by the right state $|\psi\rangle=(\psi_1,\psi_2,\ldots,\psi_n)^T$, with $q_k=\text{i}\hbar\psi_k, p_k=\bar{\phi}_k$ forming canonical variable pairs. The Hamilton's equations then take the form
\begin{equation} \label{FH}
\frac{d(\text{i}\hbar\psi_k)}{dt}=\frac{\partial\mathcal{H}}{\partial\bar{\phi}_k},  \quad
\frac{d(\bar{\phi}_k)}{dt}=-\frac{\partial\mathcal{H}}{\partial(\text{i}\hbar\psi_k)},
\end{equation}
where the Hamiltonian $\mathcal{H}$ for these canonical equations is
\begin{equation} \label{3}
\mathcal{H}(\bar{\phi}_1,\ldots,\bar{\phi}_n,\text{i}\hbar\psi_1,\ldots,\text{i}\hbar\psi_n)=\langle\bar{\phi}|h|\psi\rangle.
\end{equation}
Thus, the phase space dimension is $2n$, consistent with the canonical dynamics. The biorthogonal framework thereby allows identifying a proper phase space description despite the non-Hermiticity of $h$. When $h$ becomes Hermitian, $|\bar{\phi}\rangle$ reduces to $|\psi\rangle$, restoring the canonical structure of the standard Schr\"odinger equation.

To clarify the canonical equations, we can make a time-independent canonical transformation
\begin{equation} \nonumber
(\bar{\phi}_1,\ldots,\bar{\phi}_n,\text{i}\hbar\psi_1,\ldots,\text{i}\hbar\psi_n)\rightarrow(\bar{c}_1,\ldots,\bar{c}_n,\text{i}\hbar c_1,\ldots,\text{i}\hbar c_n),
\end{equation}
leaving the Hamiltonian $\mathcal{H}$ numerically unchanged. The canonical equations then become
\begin{equation} \label{CAN}
\frac{d(\text{i}\hbar c_j)}{dt}=\frac{\partial \mathcal{H}}{\partial \bar{c}_j},\quad\frac{d \bar{c}_j}{dt}=-\frac{\partial \mathcal{H}}{\partial(\text{i}\hbar c_j)},
\end{equation}
with $\mathcal{H}$ expressed as
\begin{equation} \label{14}
\mathcal{H}=\sum_{j=1}^n E_j\bar{c}_jc_j.
\end{equation}
The transformed variables $c_j$ and $\bar{c}_j$ evolve as Eqs.~(\ref{EV1}) and (\ref{EV2}).

The Lagrangian of the non-Hermitian dynamics can be expressed as
\begin{equation}
\mathcal{L}={\text i}\hbar\langle\bar{\phi}|\frac{\partial}{\partial t}|\psi\rangle-\langle\bar{\phi}|h|\psi\rangle,
\end{equation}
where the Lagrangian equations
\begin{equation}
\frac{d}{dt}\frac{\partial \mathcal{L}}{\partial \dot{\psi}_j}-\frac{\partial \mathcal{L}}{\partial \psi_j}=0
\end{equation}
are equivalent to the canonical equations in (\ref{FH}). This demonstrates the correspondence between the Lagrangian and Hamiltonian formulations, with the conjugate left and right states providing the necessary structure to construct a valid Lagrangian despite $h$ being non-Hermitian. The biorthonormal framework thereby allows deriving a self-consistent Lagrangian description in addition to the Hamiltonian dynamics.

\subsection{An application: recognizing the adiabatic invariants}

As an application, consider a Hamiltonian $\mathcal{H}(\mathbf{R})=\langle\bar{\phi}|h(\mathbf{R})|\psi\rangle$ that depends on slowly varying parameters $\mathbf{R}$. When the eigenenergies $E_j$ are real, according to Eqs. (\ref{6}), (\ref{relationI}) and (\ref{MOMEN}), the variables $(\bar{c}_j,c_j)$ and thus $(\bar{\phi}_j,\psi_j)$ undergo periodic oscillations. In this case, the $n$ actions
\begin{equation} \label{action1}
I_j=\frac{\text{i}\hbar}{2\pi}\oint \bar{\phi}_j d \psi_j.
\end{equation}
defined by the adiabatic theorem, are invariant. In the canonical representation of Eq. (\ref{CAN}), this becomes
\begin{equation} \label{action2}
I_j=\frac{\text{i}\hbar}{2\pi}\oint \bar{c}_j d \hbar c_j=\hbar \bar{c}_j(0)c_j(0)=\hbar|C_j|^2,
\end{equation}
showing the occupation numbers remain constant when eigenenergies are real, consistent with previous results~\cite{zhang}. This demonstrates the non-Hermitian adiabatic theorem is identical to the conventional quantum one when energies are entirely real. Adiabatic invariance holds for non-Hermitian Hamiltonians just as in standard quantum mechanics.

\subsection{An example}

As a specific example, we consider the two-level Lorentzian system from Ref.~\cite{ZhangNJP}, governed by the Bogoliubov-de Gennes equations
\begin{equation} \label{evolution}
{\text i}\hbar\frac{d}{dt}\left(\begin{array}{c}a \\b\end{array} \right)=h\left(\begin{array}{c}a \\b\end{array} \right)=
\left(\begin{array}{cc}z&x+\text{i}y\\-x+\text{i}y&-z \end{array}\right) \left(\begin{array}{c}a \\b\end{array} \right),
\end{equation}
where $x, y, z$ are real parameters. These equations describe the non-Hermitian dynamics of bosonic Bogoliubov quasiparticles in various systems.
The biorthonormal eigenstates of $h$ are
\begin{equation}  \label{eigen-vec}
|a_1\rangle=\left(\begin{array}{c}u \\v\end{array} \right), \quad |a_2\rangle=\left(\begin{array}{c}v^* \\u^*\end{array} \right), |b_1\rangle=\left(\begin{array}{c}u \\-v\end{array} \right), \quad |b_2\rangle=\left(\begin{array}{c}-v^* \\u^*\end{array} \right),\\
\end{equation}
satisfying Eqs. (\ref{right-left1}) and (\ref{right-left2}), with $|u|^2 - |v|^2 = 1$. The variables $(u,v)$ in Eq.~(\ref{eigen-vec}) are functions of the parameters $(x,y,z)$, defined explicitly as
\begin{eqnarray} \label{uv} \nonumber
&&u=-\text{sgn}(z)\frac{\left(z^2-x^2-y^2\right)^{\frac{1}{2}}+|z|}{\left\{\left[|z|+\left(z^2-x^2-y^2\right)^{\frac{1}{2}}\right]^2-x^2-y^2\right\}^{\frac{1}{2}}} \\ && v=\frac{x-\text{i}y}{\left\{\left[|z|+\left(z^2-x^2-y^2\right)^{\frac{1}{2}}\right]^2-x^2-y^2\right\}^{\frac{1}{2}}},
\end{eqnarray}
The $\text{sgn}(z)$ in Eq.~(\ref{uv}) represents the sign function of the parameter $z$. For a state $|\psi\rangle = (\psi_1,\psi_2)^T$,  if $\langle b_1|\psi\rangle \neq 0$ and $\langle b_2|\psi\rangle \neq 0$, the conjugate state is
\begin{equation}
\langle\bar{\phi}|=(\bar{\phi}_1,\bar{\phi}_2)=\frac{|C_1|^2}{\langle b_1|\psi\rangle}\langle b_1|+\frac{|C_2|^2}{\langle b_2|\psi\rangle}\langle b_2|,
\end{equation}
yielding,
\begin{eqnarray} \nonumber
\bar{\phi}_1&=&\frac{|C_1|^2u^*}{u^*\psi_1-v^*\psi_2}-\frac{|C_2|^2v}{-v\psi_1+u\psi_2}, \\
\bar{\phi}_2&=&-\frac{|C_1|^2v^*}{u^*\psi_1-v^*\psi_2}+\frac{|C_2|^2u}{-v\psi_1+u\psi_2}.
\end{eqnarray}
The Hamiltonian $\mathcal{H}(\bar{\phi}_1,\bar{\phi}_2,\text{i}\hbar\psi_1,\text{i}\hbar\psi_2)=\langle\bar{\phi}|h|\psi\rangle$ and Lagrangian $\mathcal{L}$ can then be constructed. During the evolution in Eq.~(\ref{evolution}), the overlap $\langle\bar{\phi}|\psi\rangle$ is conserved per Eq.~(\ref{conserv}).
When $z^2\geq x^2+y^2$, the two eigenenergies of $h$ are real. In this parameter region, as $(x,y,z)$ vary adiabatically, the adiabatic invariants in Eqs. (\ref{action1}) and (\ref{action2}) apply.

\section{Extension to continuous systems}

Transitioning from a discrete to a continuous system is possible when the number of degrees of freedom becomes non-countably infinite. For clarity, we adopt the coordinate representation by denoting $|\mathbf{x}\rangle$ as the coordinate eigenstate with eigenvalue $\mathbf{x}$, i.e. $\hat{x}|\mathbf{x}\rangle=\mathbf{x}|\mathbf{x}\rangle$  where $\hat{x}$ is the coordinate operator.
As in the discrete case, the eigenfunctions take the biorthonormal form of Eq.~(\ref{right-left}) but with the $j$th right and left eigenfunctions $a_j(\mathbf{x})=\langle\mathbf{x}|a_j\rangle$ and $b_j(\mathbf{x})=\langle\mathbf{x}|b_j\rangle$ now being continuous functions of $\mathbf{x}$. The biorthonormal and completeness conditions become
\begin{equation} \label{clr2}
\int b_j(\mathbf{x})^*a_k(\mathbf{x}) d\mathbf{x}=\delta_{jk}, \quad \sum_j a_j(\mathbf{x})b_j(\mathbf{x'})^* =\delta(\mathbf{x}-\mathbf{x}').
\end{equation}
In terms of the wavefunction $\psi(\mathbf{x})=\langle\mathbf{x}|\psi\rangle$, the Schr\"odinger equation is
\begin{equation} \label{NQM}
\text{i}\hbar\frac{\partial}{\partial t}\psi(\mathbf{x})=\hat{h}\psi(\mathbf{x}),
\end{equation}
where $\hat{h}$ is the Hamiltonian operator in coordinate representation, related to the matrix element $\langle \mathbf{x}'|h|\mathbf{x}\rangle=\hat{h}(\hat{x},-\text i\hbar\partial/\partial \mathbf{x})\delta(\mathbf{x}-\mathbf{x}')$.
The canonical variables follow identically to the discrete case: (i) expanding Š×(x) as
\begin{equation} \label{16}
\psi(\mathbf{x})=\sum_j c_j a_j(\mathbf{x}),
\end{equation}
(ii) constructing the canonical conjugate field
\begin{equation}  \label{7}
\bar{\phi}(\mathbf{x})=\langle\bar{\phi}|\mathbf{x}\rangle=\sum_j \bar{c}_j b_j^*(\mathbf{x}),
\end{equation}
with \begin{equation}      \label{8}
\bar{c}_jc_j=|C_j|^2.  \end{equation}
As shown, $\bar{\phi}(\mathbf{x})$ and $\psi(\mathbf{x})$ satisfy the canonical equations
\begin{equation} \label{cFH}
\frac{d(\text{i}\hbar\psi_k)}{dt}=\frac{\partial\mathbb{H}}{\partial\bar{\phi}_k}, \quad
\frac{d(\bar{\phi}_k)}{dt}=-\frac{\partial\mathbb{H}}{\partial(\text{i}\hbar\psi_k)},
\end{equation}
with Hamiltonian density
\begin{equation} \label{Hdensity}
\mathbb{H}=\bar{\phi}(\mathbf{x})\hat{h}\psi(\mathbf{x}).
\end{equation}
The Lagrangian density is
\begin{equation} \label{Ldensity}
\mathbb{L}(\mathbf{x})={\text i}\hbar\bar{\phi}(\mathbf{x})\dot{\psi}(\mathbf{x})-\bar{\phi}(\mathbf{x})\hat{h}\psi(\mathbf{x}).
\end{equation}
The equations of motion satisfy the Euler-Lagrange equation for a continuous field,
\begin{equation}
\frac{\partial\mathbb{L}}{\partial\psi}
-\nabla\cdot\frac{\partial\mathbb{L}}{\partial(\nabla\psi)}
-\frac{\partial}{\partial t}\left(\frac{\partial\mathbb{L}}{\partial\dot{\psi}}\right)=0.
\end{equation}

As an application, consider the linear non-Hermitian dynamics given by the 3D Schr\"odinger equation with a complex potential $V(\mathbf{x})$
\begin{equation} \label{Schro3}
\text{i}\hbar\frac{\partial}{\partial t}\psi(\mathbf{x},t) = -\frac{\hbar^2}{2m}\nabla^2\psi(\mathbf{x},t)+ V(\mathbf{x})\psi(\mathbf{x},t).
\end{equation}
The Lagrangian density of this system is (dropping the edge term)
\begin{equation} \label{Ldensity}
\mathbb{L}(\mathbf{x})={\text i}\hbar\bar{\phi}(\mathbf{x})\dot{\psi}(\mathbf{x})-\bar{\phi}(\mathbf{x})V\psi(\mathbf{x})
-\frac{\hbar^2}{2m}\nabla\bar{\phi}(\mathbf{x})\cdot\nabla\psi.
\end{equation}
Consider an intrinsic symmetry of the non-Hermitian dynamics
\begin{equation}  \label{sy}
\psi\rightarrow\psi'=e^{\text{i}\alpha}\psi,\quad \bar{\phi}\rightarrow\bar{\phi}'=e^{-\text{i}\alpha}\bar{\phi},
\end{equation}
with $\alpha$ a constant. This symmetry follows from the dynamics (\ref{NQM}) and biorthogonal relations (\ref{6})-(\ref{8}). For infinitesimal $\alpha$
\begin{equation}  \label{sy2}
\psi\rightarrow\psi'=\psi+\text{i}\alpha\psi,\quad \bar{\phi}\rightarrow\bar{\phi}'=\bar{\phi}-\text{i}\alpha\bar{\phi}.
\end{equation}
Applying Noether's theorem for intrinsic symmetries $j^{\mu}=\frac{\partial \mathbb{L}}{\partial (\partial_\mu \psi)} \delta\psi$ derives the conserved currents (up to an unimportant constant factor)
\begin{equation} \label{conserved}
j^{0}=\bar{\phi}\psi, \quad j^{\mathbf{x}}=\frac{i\hbar}{2m}(\bar{\phi}\nabla\psi - \psi\nabla\bar{\phi}).
\end{equation}
Since $\psi$ and $\bar{\phi}$ are intimately related, the gradients $\nabla\psi$ and $\nabla\bar{\phi}$ must be treated equally as general velocities in deriving this.

Equation (\ref{conserved}) gives,
\begin{equation}
\frac{\partial}{\partial t}(\bar{\phi}\psi) + \frac{i\hbar}{2m}\nabla\cdot(\bar{\phi}\nabla\psi - \psi\nabla\bar{\phi}) = 0
\end{equation}
This identifies a conserved charge
\begin{equation}
Q = \int \bar{\phi}(\mathbf{x})\psi(\mathbf{x}) d^3\mathbf{x}
\end{equation}
When $V(\mathbf{x})$ becomes real, $Q$ restores to the standard probability density of the Schr\"odinger wavefunction.

The conserved charge suggests that $\bar{\phi}(\mathbf{x})\psi(\mathbf{x})$ may represent the real probability amplitude of finding a non-Hermitian particle described by state $|\psi\rangle$ at position $\mathbf{x}$. Thus, by employing the Lagrangian density (\ref{Ldensity}), it should also be of considerable interest to derive the semiclassical trajectories of a wavepacket of a continuous-level non-Hermitian system, as has been done for traditional quantum systems~\cite{semi}. This would provide insight into the motion and localization properties of non-Hermitian wavepackets, which could be a potentially fruitful direction for further research.

%Overall, this highlights promising avenues to advance fundamental understanding of non-Hermitian systems.

%This indicates the non-Hermitian generalization of the Gross-Pitaevskii equation should take the self-consistent form:
%\begin{equation} \label{GP}
%i\hbar \frac{\partial}{\partial t}\psi(\mathbf{x}) = -\frac{\hbar^2}{2m}\nabla^2\psi(\mathbf{x}) + V\psi(\mathbf{x}) + g\bar{\phi}(\mathbf{x})\psi(\mathbf{x})
%\end{equation}
%where the potential $V$ is complex and $g$ characterizes the interaction strength. The conjugate state $\bar{\phi}(\mathbf{x})$, intimately dependent on $V$, can be obtained self-consistently through the outlined procedure. The Lagrangian for a specific two-mode non-Hermitian nonlinear system was given in Ref.~\cite{Liu2}. When $V$ becomes real, this reduces to the standard Gross-Pitaevskii equation.

\section{Summary}

In summary, a Hamiltonian formulation has been established for linear non-Hermitian dynamics. As applications, adiabatic invariants and a conserved charge have been obtained. All of these results naturally reduce to familiar quantities in Hermitian quantum systems. This work advances the theoretical understanding of non-Hermitian systems and highlights promising directions for future research. In particular, quantization of non-Hermitian systems and deriving semiclassical trajectories of wavepackets remain areas of considerable interest. Generalization to nonlinear non-Hermitian systems is also of great interest. Overall, this Hamiltonian formulation provides a foundation to further develop analytical tools for linear non-Hermitian systems and potentially for nonlinear ones.

\bmhead{Acknowledgments}

I thank Biao Wu from Peking University for the helpful discussion.

\noindent

\bigskip


\end{document}